\documentclass[preprint,review,12pt]{elsarticle}

\usepackage{graphicx}

\usepackage{amssymb}

\biboptions{sort&compress}

\begin{document}

\begin{frontmatter}



\title{Generation of Photon Pairs in Dispersion Shift Fibers through Spontaneous Four Wave Mixing: Influence of Self-phase Modulation}


\author{Xiaoxin Ma, Lei Yang, Xueshi Guo}
\author{Xiaoying Li \corref{cor1}}
\ead{xiaoyingli@tju.edu.cn}
\cortext[cor1]{Corresponding author. Tel.: +86-22-27406476}

\address{College of Precision Instrument and
Opto-electronics Engineering, Tianjin University, \\Key Laboratory of Optoelectronics Information Science and Technology, Ministry of Education, Tianjin, 300072,
P. R. China}

\begin{abstract}
Correlated signal and idler photon pairs with small detuning in the telecom band can be generated through spontaneous four-wave mixing in dispersion shift fibers. However, photons originated from other nonlinear processes in optical fibers, such as Raman scattering and self-phase modulation, may contaminate the photon pairs. It has been proved that photons produced by Raman scattering are the background noise of photon pairs. Here we show that photons induced by self-phase modulation of pump pulses are another origin of background noise. After studying the dependence of self-phase modulation induced photons in signal and idler bands, we demonstrate that the quantum correlation of photon pairs can be degraded by the self-phase modulation effect. The investigations are useful for characterizing and optimizing an all fiber source of photon pairs.

\end{abstract}


\end{frontmatter}



Quantum correlated photon pairs are not only the important resources for fundamental test of quantum physics, but also crucial for quantum metrology and quantum information processing. Recently, there has been growing interest in generating photon pairs by means of spontaneous four-wave mixing (SFWM) in single mode optical fibers, owing to the advantage of single spatial mode and the potential to develop into a practical quantum device.
So far, photon pairs with different kinds of wavelengths and spectra have been realized by using various kinds of optical fibers, including dispersion shift fibers (DSFs), high nonlinear optical fibers, and photonic crystal fibers ~\cite{Fiorentino02,Rarity05,Fan05a,Lim06,cohen09}. Moreover, based on these photon pairs, heralded single photon sources and entangled photon pairs with different degrees of freedom have been realized~\cite{Takesue04,TakesuePRA05,Fan08,mcmillan09,li09}.

It is worth noting that among the various kinds of photon pairs, photon pairs in the 1550 and 1310 nm bands produced via SFWM in DSFs and standard optical fibers, respectively, are promising candidates for an all-fiber source having the advantages of compact size and freedom from misalignment~\cite{Fiorentino02,Matthew09}, since off-the-shelf fiber components with high quality and low cost are commercially available in the telecom bands. Moreover, for the fiber source of photon pairs in telecom bands, which are compatible with optical network and highly desirable for long distance quantum communication, the phase-matching conditions of SFWM are satisfied when the central wavelength of pump is in the anomalous dispersion regime, and the detunings of  non-degenerate signal and idler photon pairs are usually small. Under this condition, photons having the same frequency as the individual signal or idler photons may also produced by other nonlinear processes, such as Raman scattering (RS) and self-phase modulation (SPM) of the pump pulses. It has been proved that photons produced by RS are the background noise of photon pairs. To improve the quality of the sources of photon pairs, a lot of efforts have been made on characterizing and minimizing the effect of RS, such as reducing the detuning of photon pairs, decreasing the duration of pump pulses, and lowering the temperature of optical fibers etc.~\cite{Li04,Li05c,Rarity05,TakesueOE05,Dyer08}. The influence of SPM induced photons, on the other hand, has not been fully characterized,
although there are a few experimental reports showing that SPM might become the dominate origin of scattered photons when the detuning is reduced~\cite{Li05c,Li06}.

In this paper, using 300-meter-long DSF pumped by a pulsed pump with the central wavelength in the anomalous dispersion regime, we experimentally study how the SPM of pump affect the quality of the source of photon pairs in 1550 nm band. After properly separating the SPM induced photons from those originated from SFWM and RS, we characterize its dependence upon the detuning of photon pairs, and upon the power and bandwidth of pump pulses. Moreover, we experimentally demonstrate that quantum correlation of photon pairs can be degraded by the SPM induced noise photons. Our investigations is not only helpful for developing an all fiber source of photon pairs with low background noise, but also useful for exploring its applications~\cite{Takesue07,li10}.

Our experimental setup, which is similar to that in Ref.\cite{Li06}, is shown in Fig. 1. A piece of 300 m DSF, with the zero dispersion wavelength of $\lambda
_0=1537\pm 2$\,nm is submerged in liquid nitrogen (77\,K) to suppress RS. The fiber polarization controller (FPC$_1$) is used to compensate the birefringence introduced by bending and coiling of the DSF. When the pump pulses with a
central wavelength of 1538\,nm are launched into the DSF, the parametric process of SFWM is phase-matched and the probability of the four photon scattering is significantly
enhanced. In this process, two
pump photons at frequency $\omega _{p}$ scatter through the Kerr ($\chi ^{(3)}$)
nonlinearity of the fiber to create
energy-time entangled signal and idler photons at frequencies $\omega _s$ and $\omega _i$, respectively, such that $2\omega _{p}=\omega _s+\omega
_i$. Meanwhile, accompanying SFWM, photons at frequencies $\omega _s$ and $\omega _i$ are also produced by RS and SPM of pump pulses.

The photons scattered from different nonlinear processes have their own characteristics. For SFWM, signal and idler photons are created in pair-wise and predominately co-polarized with pump, the sum of the phases of the twin photons equals to that of the two pump photons, but the phase of individual signal and idler photon is random; for RS, a signal or idler photon, originated from a pump photon scatters off an optical phonon, randomly takes the polarization and phase; while for SPM, signal and idler photons come from spectral broadening of pump pulses, will inherit the polarization and phase information of the pump photons.
If we only think of the individual signal or idler photons, those via SFWM and RS are in thermal state, while those originated from SPM are in coherent state.

To study the influence of SPM on the quantum correlation of photon pairs, it is necessary to separate the SPM induced photons from those originated from SFWM and RS. To achieve this, before launching into the DSF, pulsed pump are decomposed into horizontally
and vertically polarized components, $P_H$ and $P_V$,
respectively, by use of a polarization beam splitter PBS$_1$ (see Fig. 1). $P_H$ and $P_V$, which propagate in clockwise and counter clockwise directions, respectively, with relative phase difference of $\phi $,  will independently produce signal and idler photons via $\chi ^{(3)}$ nonlinearity.
At the output port of PBS$_1$, apart from the residual pump photons, there are three kinds of photons at frequencies $\omega _s$ and $\omega _i$ emerged: (i) co-polarized signal and idler photons created in pair-wise via SFWM; (ii) co-polarized signal and idler photons via RS; and (iii) signal and idler photons originated from SPM.
For the individual signal or idler photons produced by SPM of $P_H$ and $P_V$, respectively, the ideal first order interference is observable. Whereas the signal or idler photons produced by SFWM of $P_H$ and $P_V$, respectively, are equivalent to two independent thermal fields, it is impossible to observe the first order interference between them. So do the photons via RS.


To form the first order interference to sort out SPM induced photons, the polarization of signal and idler photons emerged at the output port of PBS$_1$ are rotated 45 degree respect to its original horizontal and vertical directions by adjusting FPC$_2$. Then, the photons propagate through PBS$_2$ and filter F. The filter F has the function of well separating signal, idler, and pump photons, and the central wavelengthes of the detected signal and idler photons with a FWHM of about 0.65 nm can be varied by adjusting F, which is realized by cascading a free-space double-grating spectral filter with tunable filters~\cite{Li06}. Since the Kerr nonlinearity is weak, only about 0.1 photon pairs are typically produced by a $5$ ps pump pulse containing $5\times 10^{7}$ photons, to reliably detect the photon pairs, the pump-rejection ratio provided by F should be greater than 100 dB. When the signal and idler photons are detected by using single photon detectors SPD$_3$ and SPD$_2$, respectively, and the powers of $P_H$ and $P_V$ are equal, the measured counting rate of SPD$_{2(3)}$ can be expressed as:
\begin{equation}\label{1}
   N_{t} = N_{F}  + N_{R}  + N_{S} \left( {1 + \cos \phi } \right)
\end{equation}
where $N_{F}$, $N_{R}$ and $N_{S}$ are count rate of photons produced from SFWM, RS and SPM, respectively. By scanning the relative phase $\phi$ between P$_H$ and P$_V$, a cosine variation can be obtained, from which $N_S$ can be extracted. Since the counting rate of SPM induced photons in signal band is slightly smaller than that in idler band due to asymmetry in spectrum broadening of the pump, for the counting measurement of individual signal or idler photons, we only present the results of idler photons for convenience.

The function of scanning the relative phase difference $\phi$ is realized by decompounding the linearly polarized pump in the port labelled "pump in" and adding separate
free-space propagation paths for the two decomposed pumps, $P_H$ and $P_V$, with use of
the PBS$_3$, quarter-wave plates
 QWP$_1$ and QWP$_2$, and mirrors M$_1$ and M$_2$ (see Fig. 1). This arrangement is equivalent to a polarization interferometer formed between PBS$_3$ and PBS$_2$.
The linearly polarized pump with a Gaussian shaped spectrum is obtained by using a filter to spectrally carve the output of a mode-locked femto-second fiber laser having a repetition rate of about 41 MHz~\cite{li10}, and the time bandwidth product of pump is about 0.6.
M$_2$ is mounted on a piezoelectric-transducer (PZT)-driven
translation stage, which allows precise adjustment of the
relative delay and phase difference between $P_H$ and $P_V$. In our experiment, the relative delay is adjusted to be $0$. To monitor the phase difference $\phi$, the residual pump photons, separated by the filter F, are attenuated to single photon level and then measured by using SPD$_1$. The power of $P_H$ is adjusted to be equal to that of $P_V$ by properly regulating half-wave plate HWP$_1$. Therefore, the counting rate of SPD$_1$, $n_p$, can be expressed as
\begin{equation}\label{1}
   n_p \propto  N_p \left( {1 + \cos \phi } \right),
\end{equation}
where $N_p$ is determined by the intensity of P$_H$ (P$_V$).

The SPDs used for conducting the photon counting measurement are InGaAs/InP avalanche photodiode based SPDs (PLI-AGD-SC-Rx and id200) operated in the gated-Geiger mode. The 2.5-ns-wide gate pulses arrive at a rate of 1.29 MHz, which is 1/32 of the repetition rate of the pump pulses, and the dead time of the gate
is set to be 10 $\mu$s. The timing of gate pulses is adjusted by a digital delay generator to coincide with the arrival of idler, signal, or attenuated pump photons. The electrical signals produced by the SPDs in response to the incoming photons (and dark events), reshaped into 100-ns wide
TTL pulses, are then
acquired by a computer-controlled analog-to-digital (A/D)
board (National Instrument, PCI-6251). Thus, both the single counts of each SPD
and two-fold coincidences of SPD$_2$ and SPD$_3$ acquired from
different time bins can be determined because the A/D card records
all counting events. The total detection efficiency of the idler (signal) photons is about $2\% $.

Before characterizing the SPM induced photons in idler band, we first find out the possible smallest value of detuning between idler (signal) and pump photons, $\Omega =(\omega _{i(s)}-\omega _p)/2\pi $, at which the signal and idler photons contained in the incident pump pulses, P$_H$ and P$_V$, are negligible, and pump-rejection ratio provided by filter F in signal and idler bands is greater than 100 dB. In the measurement, the FWHM of the Gaussian shaped pump pulses is about 0.95 nm, and the DSF is taken away. The photon counting measurements in idler band are conducted when the detuning $\Omega$ is varied with a step of 50 GHz (0.4 nm). Meanwhile, to monitor the phase difference $\phi$, the counting rate of the attenuated pump photons $n_p$ is also recorded.
At each detuning, we record the single counts of SPD$_1$ and SPD$_2$, respectively, as the relative phase difference $\phi$ is scanned. Figure 2(a) plots the measured counting rates as a function of $\phi$ when the average pump powers of both P$_H$ and P$_V$ are 0.1 mW ($\sim 2\times 10^{7}$ photons/pulse). It is clear that the count rate presents periodic variation at the detuning $\Omega=0.4$ THz (3.2 nm), showing there are photons detected in idler (and signal) band even if nonlinear medium DSF is absent; whereas the periodic variation is not obvious when $\Omega$ is greater than 0.45 THz (3.6 nm). Taking the total detection efficiency in idler band into account, we are able to deduce that the ratio between the number of photons in idler and in pump bands per pulse is about $2.5\times 10^{-10}$ and $2.5\times 10^{-11}$ for $\Omega=0.4$ THz and $\Omega=0.45$ THz, respectively. Therefore, the pump-rejection ratio of filter F is more than 100 dB for $\Omega$ greater than 0.45 THz (3.6 nm), namely, the possible smallest detuning is about 3.6 nm.

Considering the spectra of pumps, P$_H$ and P$_V$, might be broadened in DSF due to SPM, we start to characterize the SPM induced photons at the detuning of 4 nm, which is slightly greater than the smallest detuning deduced from the above measurement. In this measurement, we put the DSF back and make the same measurement again when the average powers for both P$_H$ and P$_V$ are about 90 $\mu$W. Figure 2(b) shows the counting rates recorded by SPD$_1$ and SPD$_2$, respectively, as a function the relative phase difference $\phi$. Fitting the data with Eqs. (1) and (2), we find that the periodicity of the count rate in idler band is the same as that of the pump, and the extracted rate $N_{S} $, proportional to the SPM induced photons, is about $70\%$ of the total counting rate $N_t^{\prime} = N_{F}  + N_{R}  + N_{S}$. Therefore, the photons originated from SPM effect could almost dominate over SFWM even if the filter F can provide a pump isolation in excess of 100 dB.

To figure out the power dependence of $N_{S} $, we further increase the detuning to 4.4 nm, and repeat the above measurement at different power levels of P$_H$ (P$_V$). At each power level, after extracting the the counting rate $N_{S}$ from the measured results of $N_t$, we fit the counting rate of the photons originated from RS and SFWM and, $N_{R}  + N_{F}$, with the second-order polynomial $s_1 P_{ave}  + s_2 P_{ave}^2 $, where the linear and quadratic coefficients, $s_1$ and $s_2$, respectively determine the strengths of RS and SFWM in DSF, and $P_{ave}$ refers to the average power of P$_H$ (P$_V$). The main plot in Fig. 3 shows the counting rates $N_{S} $ and $N_F  + N_R$, respectively, as a function of the average power of pump P$_H$ (P$_V$). The inset of Fig. 3
sketches the ratio  $N_{S} /N_{F} $ versus the average power of P$_H$ (P$_V$). One sees that $N_{S}$ exponentially increases with the increase of pump power. Additionally, although $N_{S}$
is small at low pump power levels, it is still comparable with the photon counting rate $N_R$ originated RS.

The experimental results in Fig. 2 and Fig. 3 indicate that without carefully excluding the noise photons coming from SPM, it would be impossible to precisely characterize the photon pairs produced by SFWM. Therefore, it is necessary to further investigate the factors influencing the counting rate $N_{S} $.
Firstly, we study the influence of detuning at different pump powers. The measurement process is similar to previous, except the detuning is adjusted from 4.4 to 6 nm. At each detuning, we deduce the ratio $N_{S} /N_{F} $ when the power level of P$_H$ (P$_V$) is 140, 190 and 240 $\mu$W, respectively, as shown in Fig. 4. One sees that at a certain pump power, $N_{S} /N_{F} $ decreases rapidly with the increase of detuning; while at a certain detuning, $N_{S} /N_{F} $ increases with the increase of pump power. For the power level of $P_H=P_V=190$ $\mu$W, $N_{S} /N_{F} $ is less than $5\%$ when detuning is greater than 4.8 nm; while for the power level of $P_H=P_V=240$ $\mu$W, to obtain the ratio $N_{S} /N_{F} $ less than $5\%$, the detuning should be greater than 5.2 nm. The result indicates that in order to suppress the SPM effect, the higher the power of pump, the greater the detuning is required.

Secondly, we study the influence of the bandwidth of pump. After modifying the FWHM of the pump to 0.65 nm, we make the photon counting measurement again by varying the detuning from 4 to 4.8 nm for the pumps with power level of $P_H=P_V=240$ $\mu$W. The results are also plot in Fig. 4. One sees that for the same detuning, the ratio $N_{S} /N_{F} $ obtained by using the pump with FWHM of 0.65 nm is the smallest. Moreover, if we consider that the counting rate of photons via SFWM, $N_F$, obtained by using this pump is higher than that obtained by using the pump with FWHM and power of 0.95 nm and 140 $\mu$W~\cite{li10}, respectively, it is then obvious that the negative effect of SPM to the fiber source of photon pairs decreases with the reduction of the bandwidth of pump.

In fact, the experimental results displayed from Fig. 2 to Fig. 4 can be qualitatively explained by using a simplified theoretical model. Assuming the strong pumps launching into DSF are transform limited, the Fourier transform of the pulses propagating through the DSF is given by
\begin{equation}
\label{spectrum}
    E(z,\omega )\propto \int \sqrt{P_p}\exp (-\frac{T^2}{2T_0^2})\exp [i\gamma P_pze^{ -T^2/T_0^2}]\exp [i(\omega -\omega _{p0} ) T]dT.
\end{equation}
where $P_p$ and $\omega _{p0} $ are the peak power and the central frequency of pump, respectively; $\gamma$ is the nonlinear interaction coefficient. In this case, the number of photons in the original pump pulse can be expressed as
\begin{equation}
    N_p  = \frac{1}{{2\pi \hbar \omega _{p0} }}\int_{ - \infty }^\infty  {\left| {E(0,\omega )} \right|^2 d\omega },
\end{equation}
and the number of the SPM induced photons in idler/signal band can be written as
\begin{equation}\label{spmnumber}
   N_{Si(s)}  = \frac{1}{{2\pi \hbar \omega _{i(s)0} }}\int_{ - \infty }^\infty  {\left| { E(L,\omega )} \right|^2 f_{i(s)} (\omega )d\omega },
\end{equation}
where $L$ is the length of DSF, $f_{i(s)} (\omega )$ and $\omega _{i(s)0} $ are the transmission function and central frequency of the filter in idler (signal) band, respectively.

Because the detuning of photon pairs produced via SFWM in DSF is small, to ensure a reliable detection of photon pairs, we need to take the rejection of the pump photons with spectrum broadened by SPM into account. Therefore, for the filter F in signal and idler band, not only a greater than 100 dB isolation to the original pump photons is necessary, the condition
\begin{equation}\label{rejection}
  N_{Si(s)}/N_p <10^{ - 10}
\end{equation}
is also required. Assuming the group velocity dispersion effect can be neglected, at the output port of the DSF, the spectral broadening factor of the pump pulse is about $\sqrt {1 + (0.88\gamma P_p L)^2 }$~\cite{Agrawal}. In this case, for the filter in idler (signal) band described by a Gaussian function, the minimum detuning satisfying Eq. ({\ref{rejection}) is approximately:
\begin{equation}\label{detuning}
   \left| {\lambda _{i(s)0}  - \lambda _{p0} } \right| > \sqrt {10\ln 10} \sqrt {\sigma _p^2 [1 + (0.88\gamma P_p L)^2 ]  + \sigma _{i(s)}^2 }
\end{equation}
where $\lambda _{p(i,s)0}$ is the central wavelength of the pump (idler, signal) photons; $\sigma _p=\lambda _{p0}^{2}/(2\pi c T_0)$ is 1/e half-width of input pump, and $ \sigma _{i(s)}$ is 1/e half-width of filter in idler (signal) band. According to Eq. (\ref{detuning}) we plot the relationship between the minimum detuning and average pump power for the pump with FWHM of 0.95nm and 0.65nm, respectively. It is clear that the tendency shown in Fig. 5 agrees with our experimental data, except the calculated minimum detuning is smaller than that in our experiments. We think the main reason accounting for this departure is: the pump pulses are also influenced by chromatic dispersion, so the spectrum broadening is more severe. If the chromatic dispersion parameters of the DSF can be accurately measured, it would be possible to obtain a rigorous result of power distribution of the broadened pump in the spectral domain by using time-domain Schindinger equation~\cite{Agrawal}.

Having investigated the dependence of $N_{S}$, we finally demonstrate that photon pairs via SFWM can be contaminated by the SPM induced photons. In this experiment, we rotate HWP$_1$ to ensure only the vertically polarized pump $P_V$ is launched into DSF ($P_H=0$), and adjust FPC$_2$ to select the co-polarized signal and idler photons (see Fig. 1).  During the measurement, the FWHM of pump is about 0.95 nm, the coincidence and accidental coincidence rate of signal and idler photons $C_c$ and $C_a$, produced by the same pulse and adjacent pulses, respectively, are recorded at different power levels. To show the influence of SPM, the measurement is conducted for the detunings of 4.4 and 5.6 nm, respectively. Moreover, for each measurement, we calculate the true coincidence by subtracting $C_a$ from $C_c$, which is a reflection of the quantum correlation of photon pairs. The main plot in Fig. 6 shows the true-coincidence to accidental-coincidence ratio (TAR) $R_T=\frac{C_c-C_a}{C_a}$ as a function of pump powers. One sees that when the pump power is lower than 0.15 mW, TAR obtained with detuning of 4.4 nm is slightly higher. However, when the pump power is higher than 0.2 mW, TAR obtained with detuning of 4.4 nm is obviously smaller, and the difference of TAR obtained under the two kinds of detunings increases with the increase of pump.

The results in the main plot of Fig.6 can be understood by analyzing the dependence of TAR $R_T$. At a certain pump power, the numerator of $R_T$--the true coincidence originated from photon pairs, are the same for the two kinds of detunings because of the broadband nature of SFWM in DSF. However, the denominators of $R_T$, which is the accidental coincidence $C_a$ determined by the total counts in signal and idler bands, including photons via the nonlinear processes of SFWM, RS and SPM, depends on detunings. At a certain pump power, single counts originated from SFWM are the same for the two cases, while the single counts contributed by RS are lower for the case with smaller detunings. Therefore, if SPM induced photons are negligibly small, the accidental coincidence rate for detuning of 4.4 nm should be lower than that for detuning of 5.6 nm. This is why TAR obtained with detuning of 4.4 nm is higher in the range of lower pump power levels. However, with the increase of pump power, the accidental coincidence for detuning of 4.4 nm grows faster than that for detuning of 5.6 nm, because SPM induced photons with smaller detuning increases more rapidly (see Fig. 4). Therefore, at high pump power levels, TAR obtained with the detuning of 4.4 nm become smaller than that with detuning of 5.6 nm, showing SPM induced photons are the noise background of photon pairs.

To further illustrate that the SPM induced photons are the background noise of photon pairs, we plot true coincidence rate as a function of the single count rate in idler band $N_{t}^{\prime}$, as shown in the inset of Fig. 6. One sees that there is no much difference between the data obtained with different detunings for $N_{t}^{\prime}$ in the low rate region. However, the true coincidence rate obtained with detuning of 4.4 nm is smaller for $N_{t}^{\prime}$ in the high rate region, within which the portion of SPM induced photons is higher for the data obtained with a smaller detuning. This is because SPM induced photons in signal and idler bands are not created in pair, thus, does not contribute to the true coincidence rate.

In conclusion, using 300 m DSF pumped by a pulsed pump with central wavelength in the anomalous dispersion regime of DSF, we have not only studied the influence of the SPM induced photons by measuring its dependence upon the detuning, upon the bandwidth and average power of pump pulses, but also clearly demonstrated that photon pairs via SFWM can be contaminated by the SPM induced noise photons.

Generally speaking, to develop an all fiber source of photon pairs in the telecom band with high quality, the side effect of SPM can be eliminated or mitigated by increasing the detuning, by decreasing the bandwidth of pump, or by exploiting a Sagnac fiber loop which has the ability to provide an extra 30 dB rejection to the spectrally broadened pump pulses~\cite{Li04,Mortimore88}. However,increasing detuning or decreasing the bandwidth of pump will result in the increased background noise via RS~\cite{Li05c,Rarity05}. Therefore, to improve the quality of the source of photon pairs, one need to balance the effects of RS and SPM by optimizing the source parameters, including the detuning and the bandwidth of pump, so that the total noise photons produced by RS and SPM are maximally suppressed. Our detailed investigations of the influence of SPM is not only helpful for developing an all fiber source of photon pairs with low background noise in the telecom band, but also useful for exploring its applications.

\section*{Acknowledgement}
This work was supported in part by the NSF of China
(Grant No. 10774111, 11074186), the Specialized Research Fund for the Doctoral Program of Higher Education of China (Grant No. 20070056084), 111 Project (Grant No. B07014), and the State Key
Development Program for Basic Research of China (Grant No. 2010CB923101).



\begin{figure}[htb]
\centering
\includegraphics[width=9cm]{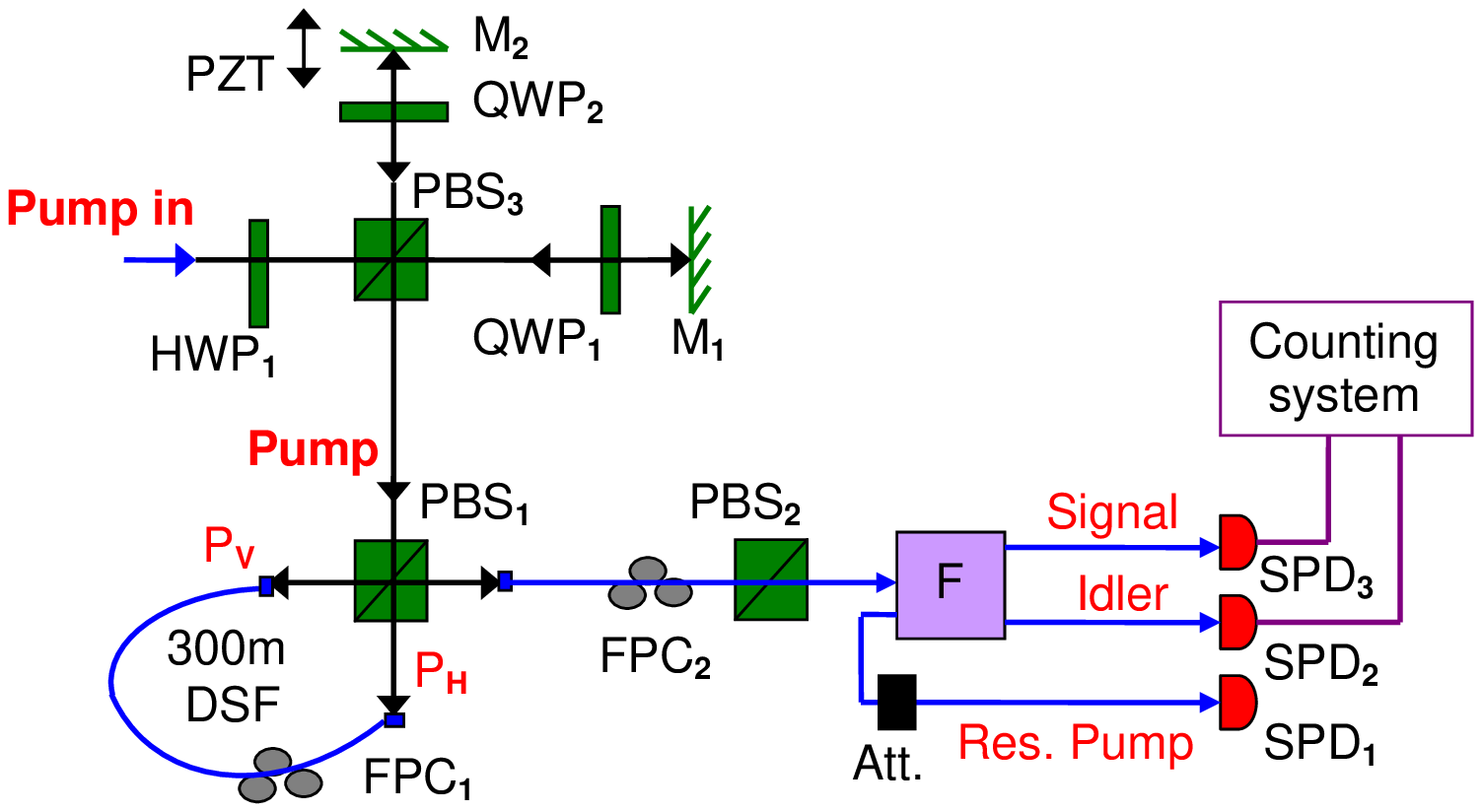}
\caption{(Color online) A schematic of the experimental setup. PBS: polarization beam splitter; QWP: quarter-wave plate; HWP: half-wave plate; M: mirror; PZT: piezoelectric-transducer; FPC: fiber polarization controller; F: filter; Att.: attenuator; SPD: single photon detector.
}
\end{figure}

\begin{figure}[htb]
\centering
\includegraphics[width=9cm]{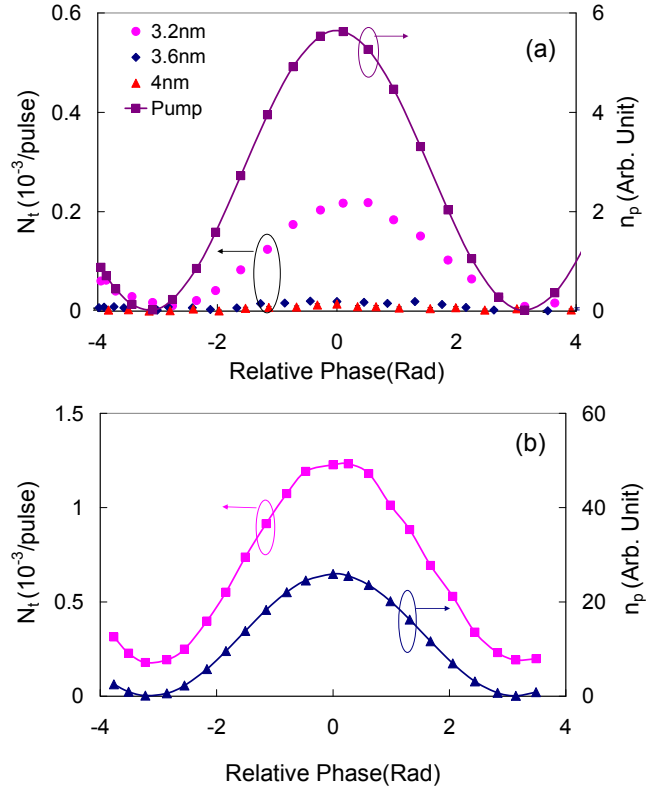}
\caption{(Color online) Measured counting rate of the idler photons and attenuated pump photons, $N_{t} $ and $n_{p} $, as a function of relative phase $\phi$ (a) when DSF is taken out, and the detuning is 3.2, 3.6 and 4 nm, respectively; and (b) when DSF is put back, and the detuning is 4 nm.
}
\end{figure}

\begin{figure}[htb]
\centering
\includegraphics[width=8cm]{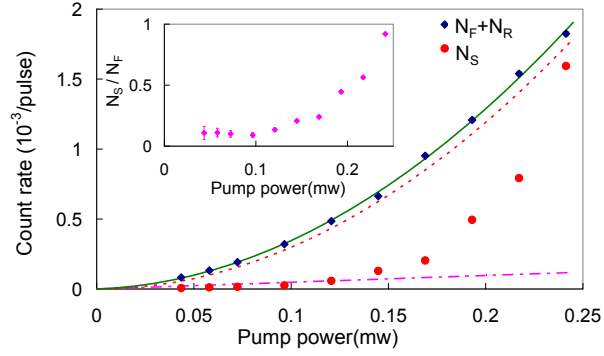}
\caption{(Color online) The deduced counting rate of photons produced by SPM, $N_{S}$, and the sum of SFWM and RS, $N_F  + N_R$, respectively, as a function of the average power of pump P$_H$ (P$_V$). A second-order polynomial $N_{R}  + N_{F}  = s_1 P_{ave}  + s_2 P_{ave}^2 $ is used to fit the deduced data, and the contributions of linear scattering $N_{R}= s_1 P_{ave}$(dash-dot line) and quadratic scattering $ N_{F}  = s_2 P_{ave}^2 $ (dot line) are plotted separately. The inset is the ratio $N_{S} /N_{F} $, as a function of the average power of P$_H$ (P$_V$). The measurement is done at the detuning 4.4 nm.
}
\end{figure}

\begin{figure}[htb]
\centering
\includegraphics[width=8cm]{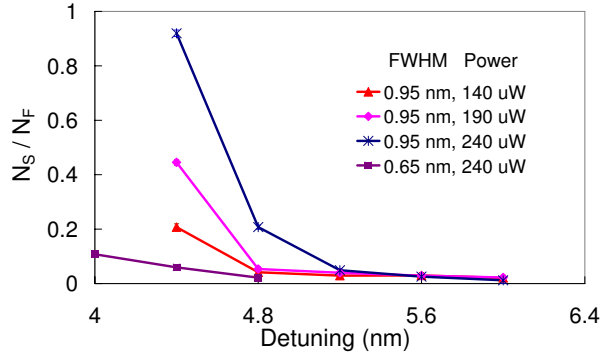}
\caption{(Color online) The ratio of photons originated from SPM and SFWM, $N_{S} /N_{F}$, as a function of the detuning for the pump with different power levels and with different bandwidths. The solid lines are only for guiding eyes.
}
\end{figure}

\begin{figure}[htb]
\centering
\includegraphics[width=8cm]{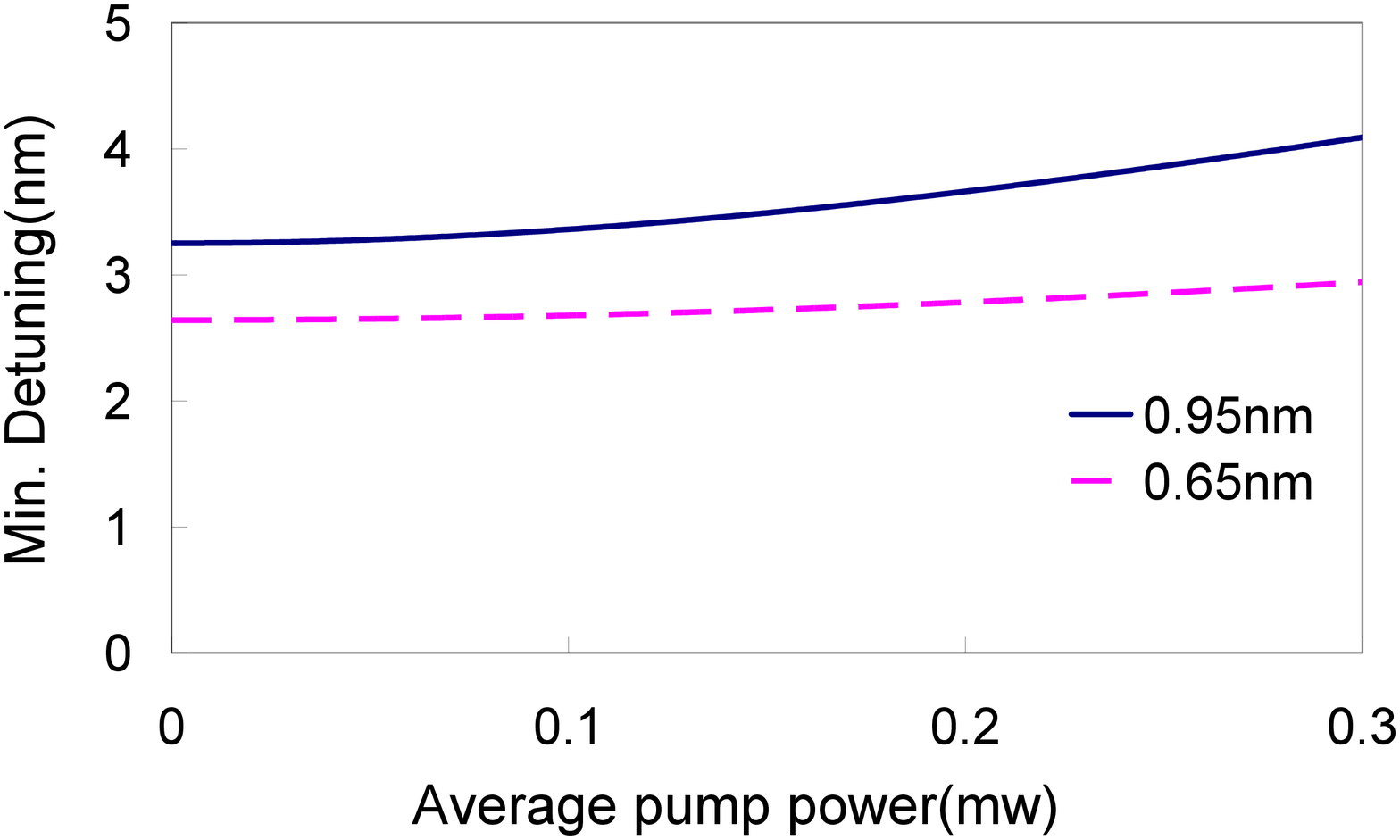}
\caption{(Color online) Calculated minimum detuning of signal and idler photon pairs versus the average power of pump pulses with different FWHM.
}
\end{figure}

\begin{figure}[htb]
\centering
\includegraphics[width=8cm]{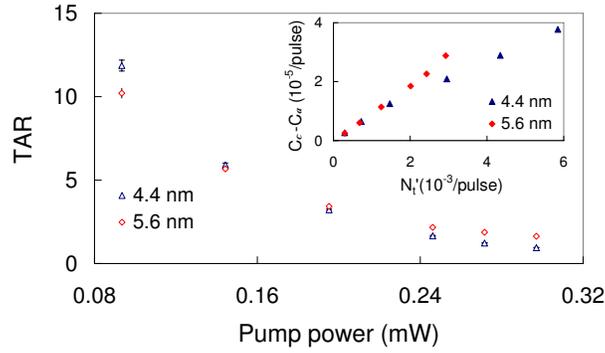}
\caption{(Color online) The true-coincidence to accidental-coincidence ratio (TAR) of photon pairs versus pump power for detuning of 4.4 nm and 5.6 nm, respectively. The inset shows true coincidence rate $C_c-C_a$ as a function of the single count rate in idler band $N_{t}^{\prime}$.
}
\end{figure}

\end{document}